\RequirePackage{fix-cm}
\documentclass[smallextended]{svjour3}       
\smartqed  

\usepackage{graphicx}
\usepackage{amsmath,amssymb}
\usepackage[numbers]{natbib}

\begin{document}

\title{Pyrolytic Graphite Sheet\thanks{This work was financially supported by Grant-in-Aid for Scientific Research (B) (Grant No.~15H03684), and Challenging Exploratory Research (Grant No.~15K13398) from JSPS.}
}
\subtitle{a New Adsorption Substrate for Superfluid Thin Films}


\author{Sachiko Nakamura$^1$ \and Daisuke Miyafuji$^2$ \and Ryo Toda$^1$ \and Tomohiro Matsui$^2$ \and Hiroshi Fukuyama$^{1,2}$
}


\institute{S. Nakamura \at \email{snakamura@crc.u-tokyo.ac.jp} \at H. Fukuyama \at \email{hiroshi@phys.s.u-tokyo.ac.jp} \at
$^1$	Cryogenic Research Center, the University of Tokyo, 2-11-16, Yayoi, Bunkyo-ku, Tokyo 133-0032, Japan \\         
$^2$	Department of Physics, the University of Tokyo, 7-3-1, Hongo, Bunkyo-ku, Tokyo 133-0033, Japan
}

\date{Received: date / Accepted: date}

\maketitle

\begin{abstract}
We have measured surface morphology and gas adsorption characteristics of uncompressed pyrolytic graphite sheet (uPGS) which is a candidate substrate for AC and DC superflow experiments on monolayers of $^4$He below $T= 1$\,K.
The PGS is a mass-produced thin graphite sheet with various thicknesses between 10 and 100\,$\mu$m.
We employed a variety of measuring techniques such as imagings with optical microscope, SEM and STM, Raman spectroscopy, and adsorption isotherm.
PGS has smooth and atomically-flat external surfaces with high crystallinity. 
Although the specific surface area ($\leq 0.1$\,m$^2$/g) is rather small, by making use of its smooth external surface, the thinnest uPGS of 10\,$\mu$m thick is found to be suitable for the superflow experiments on the strictly two-dimensional helium systems.
\end{abstract}

\section{Introduction}
\label{intro}
A few atomic layers of helium (He) adsorbed on graphite substrate are experimental realization of correlated two-dimensional (2D) quantum systems. 
New types of superfluidity, which should not be a simple extrapolation of thin superfluid $^4$He films of several tens atomic layers thick, are expected there. 
Crowell and Reppy~\cite{PhysRevLett.70.3291} found a novel superfluidity below 400\,mK, which is reentrant as a function of density, in the second monolayer of $^4$He adsorbed on Grafoil~\cite{Grafoil}, a flexible exfoliated graphite sheet, using the torsional oscillator technique. 
Recently, similar observations followed using the same method and the same type of substrate~\cite{1742-6596-150-3-032096, nyeki2017}. 
Among them, Nyeki et al.~\cite{nyeki2017} reported complicated density and temperature evolutions of the superfluid response in detail. 
Near the corresponding density region, the existence of the quantum liquid crystal phase, a new class of matter where spatial symmetry is partially broken keeping finite fluidity even at $T = 0$, has been proposed by the heat capacity measurement with a ZYX exfoliated graphite~\cite{PhysRevB.94.180501}. 
Thus the observed reentrant superfluidity~\cite{PhysRevLett.70.3291,1742-6596-150-3-032096, nyeki2017} could be a novel one where spatial and superfluid orders are intertwined, attracting great attention as a possible experimental realization of supersolidity in 2D.
Similar possibilities are discussed in different 2D quantum systems such as cold molecule gases~\cite{Hazzard2017}.

However, superfluid detection efficiencies ($\eta \equiv 1- \chi$) in those torsional oscillator experiments~\cite{PhysRevLett.70.3291,1742-6596-150-3-032096, nyeki2017} are rather poor (only $\eta = 0.02$--$0.06$) because of complicated structure and discontinuity of the Grafoil surface. 
Here $\chi$ is the tortuosity factor ($= 0.94$--$0.98$).
Such structures are inevitably created during the production process, i.e., powdering natural flake graphite, exfoliating the powders at high temperature, and rolling them into a sheet under high pressure. 
This arises a doubt that liquid or amorphous $^4$He localized in the substrate heterogeneities might be responsible for the detected superfluid responses.
Therefore, it is crucial to develop a more efficient graphite substrate for detection of the AC or DC superflow in monolayers of $^4$He.

From the previous heat capacity measurements~\cite{PhysRevB.94.180501,PhysRevLett.38.501,qfs2013}, ZYX was proven to be a better substrate than Grafoil for studies of phase transitions in 2D helium systems because of its much larger single crystallite (platelet) size and hence much less heterogeneities.
However, we have found that the surface of ZYX is covered by many deep crevices created during the exfoliation process, which may interrupt superflow in adsorbed monolayer of $^4$He.
This problem seems to be relevant in our preliminary torsional oscillator experiment as described in Section~2.

Considering the above situation, in this article, we propose the pyrolytic graphite sheet (PGS)~\cite{pgs} as a suitable graphite substrate for detection of superfluidity in monolayers of $^4$He.
PGS is a very thin graphite sheet of 10--100\,$\mu$m in thickness with a highly-oriented single-crystal-like structure made from stacked thin polymer films. 
We will make use of the \textit{external} surface for helium adsorption unlike the exfoliated graphites like Grafoil and ZYX. 
Results of extensive characterizations of its surface structure and crystallinity are given here with comparisons to results for Grafoil and ZYX.
The characterizations include (i) surface morphology observations with optical microscope, scanning electron microscope, and scanning tunneling microscope, (ii) crystallographic obsevation with Raman scattering, and (iii) adsorption isotherms.
Other properties of PGS such as thermal conductivity, electrical conductivity and crystal analysis have already been published elsewhere~\cite{NAKAMURA2017118}.

\section{Morphology}
\subsection{Observations with optical and scanning electron microscopes}
\begin{figure}[htbp]
\includegraphics[width=0.97\columnwidth]{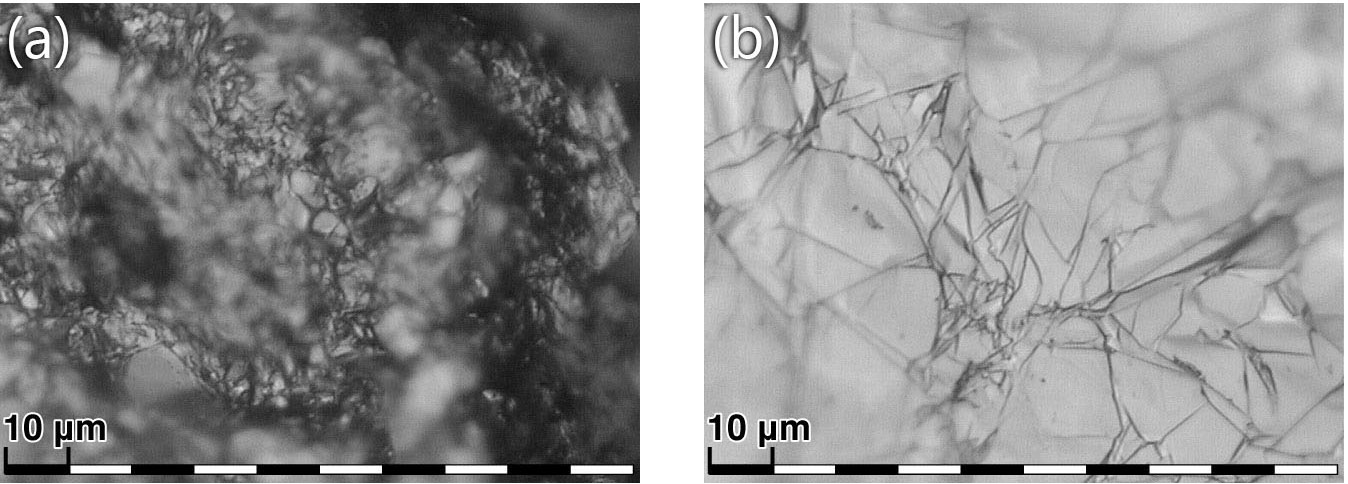}
\caption{Optical microscope images of cleaved surfaces of (a) Grafoil and (b) ZYX.}
\label{photo-GrafoilZYX}
\end{figure}

Morphology of cleaved surfaces of three different kinds of graphite substrates was checked in a relatively macroscopic scale with an optical microscope~\cite{vh5000}.
Before showing results of PGS, we first discuss those of Grafoil and ZYX. 
As shown in Fig.~\ref{photo-GrafoilZYX}(a), the surface of Grafoil is quite irregular with a mosaic structure consisting of plates of random size. 
As shown in Fig.~\ref{photo-GrafoilZYX}(b), the surface of ZYX is much more uniform.
However, we found that it is fully covered by a network of deep crevices of 1--10~$\mu$m long and that similar structures are seen at any cleaved surfaces.
The crevices must be created by the exfoliation process of this product.
The structure would prevent simply connected superflow in adsorbed $^4$He films.
This could be the reason for a rather extended superfluid onset observed in a bilayer of liquid $^4$He adsorbed on ZYX in our preliminary torsional oscillator experiment (see Appendix~A).

\begin{figure}[htbp]
\includegraphics[width=0.97\columnwidth]{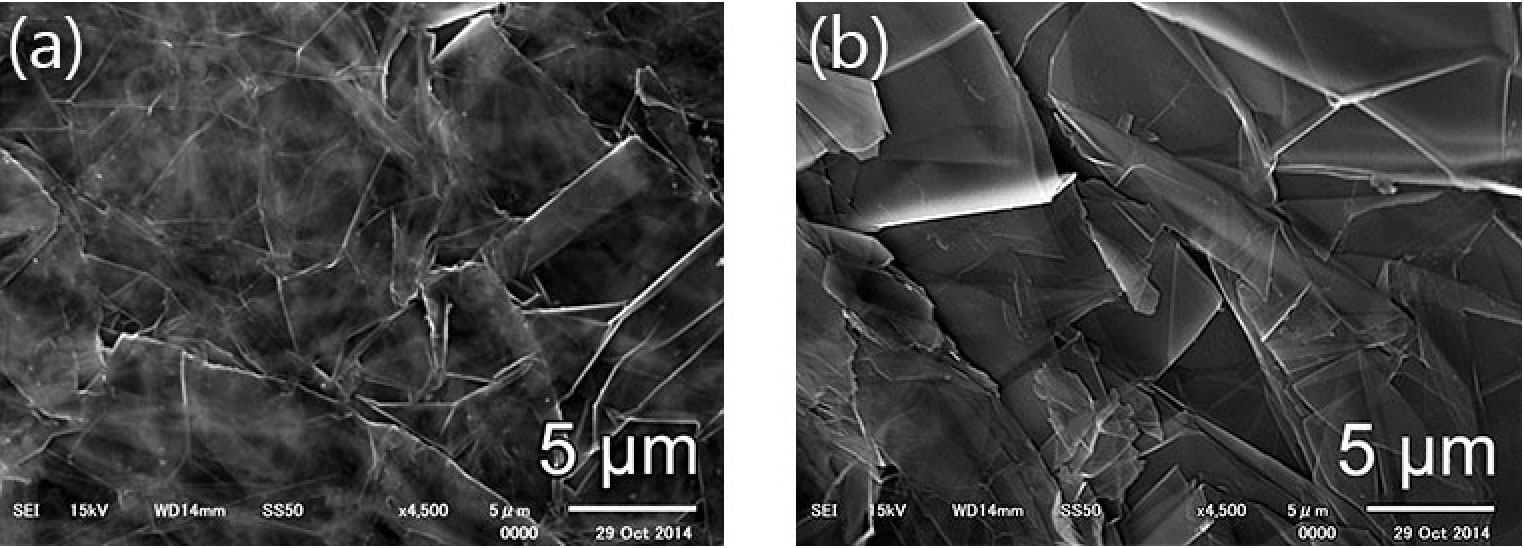}
\caption{Secondary electron SEM images of cleaved surfaces of (a) Grafoil and (b) ZYX.}
\label{SEM-GrafoilZYX}
\end{figure}

Secondary electron imaging with a scanning electron microscope (SEM)~\cite{sem} is more sensitive to a height variation of surface than the optical microscope imaging.
Figures~\ref{SEM-GrafoilZYX}(a) and (b) are such images of cleaved surfaces of Grafoil and ZYX. 
They have a piled-up structure consisting of 1--20\,$\mu$m wide patches with sharp edges. 
This structure presumably reflects small graphite flakes, a raw material, in the case of Grafoil and a trace of rupture during exfoliation in the case of ZYX, respectively. 
In any case, it will largely reduce the connectivity of superflow in adsorbed $^4$He thin films as was actually found in the previous torsional oscillator experiments using Grafoil~\cite{PhysRevLett.70.3291,1742-6596-150-3-032096,nyeki2017}.

Next, we will show results for PGS.
Two different kinds of PGS have been examined in this work.
One is uncompressed PGS (uPGS), and the other is commercial PGS (cPGS)~\cite{pgs,NAKAMURA2017118}.
uPGS is brittle and inflexible, while cPGS, which is produced by compressing uPGS, is flexible like paper.
Note that numbers denoted to specify the sample thickness below are nominal ones.
Actual thicknesses measured by micrometer are 13$\pm$2\,$\mu$m for cPGS-10$\mu$m, 19$\pm$2\,$\mu$m for uPGS-10$\mu$m, 29.7$\pm$0.6\,$\mu$m for uPGS-17$\mu$m, 56$\pm$3\,$\mu$m for uPGS-25$\mu$m, and 145$\pm$4\,$\mu$m for uPGS-100$\mu$m.

\begin{figure}[htbp]
\includegraphics[width=0.97\columnwidth]{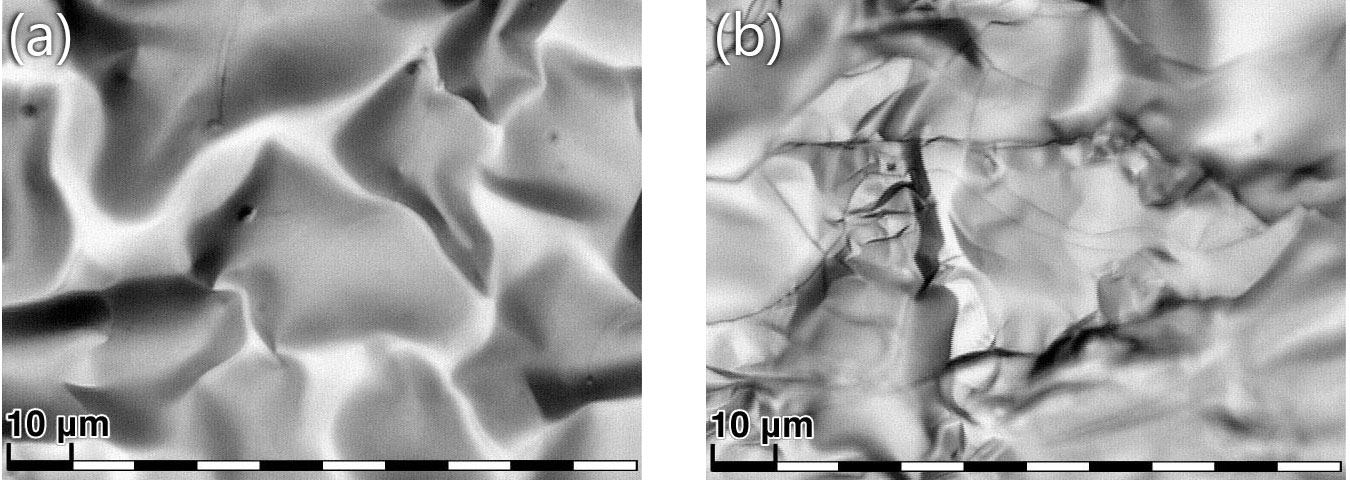}
\caption{Optical microscope images of cleaved surfaces of (a)uPGS-10$\mu$m and (b) cPGS-10$\mu$m.}
\label{photo-uPGScPGS}
\end{figure}

The surface morphology of PGS is largely different from those of Grafoil and ZYX. 
Figure~\ref{photo-uPGScPGS}(a) is an optical microscope image of a cleaved surface of uPGS-10$\mu$m.
The surface is quite smooth with a winkle (or wavy) structure of the length scale of 20--40\,$\mu$m without any crevices.
On the other hand, in the case of cPGS-10$\mu$m (Fig.~\ref{photo-uPGScPGS}(b)), the surface is little more complicated with small structures as well as sharp creases which must be created by the final compression process.
Thus cPGS must be less suitable for superflow experiments than uPGS, and we will concentrate mostly on characterization of uPGS hereafter.

\begin{figure}[htbp]
\includegraphics[width=0.97\columnwidth]{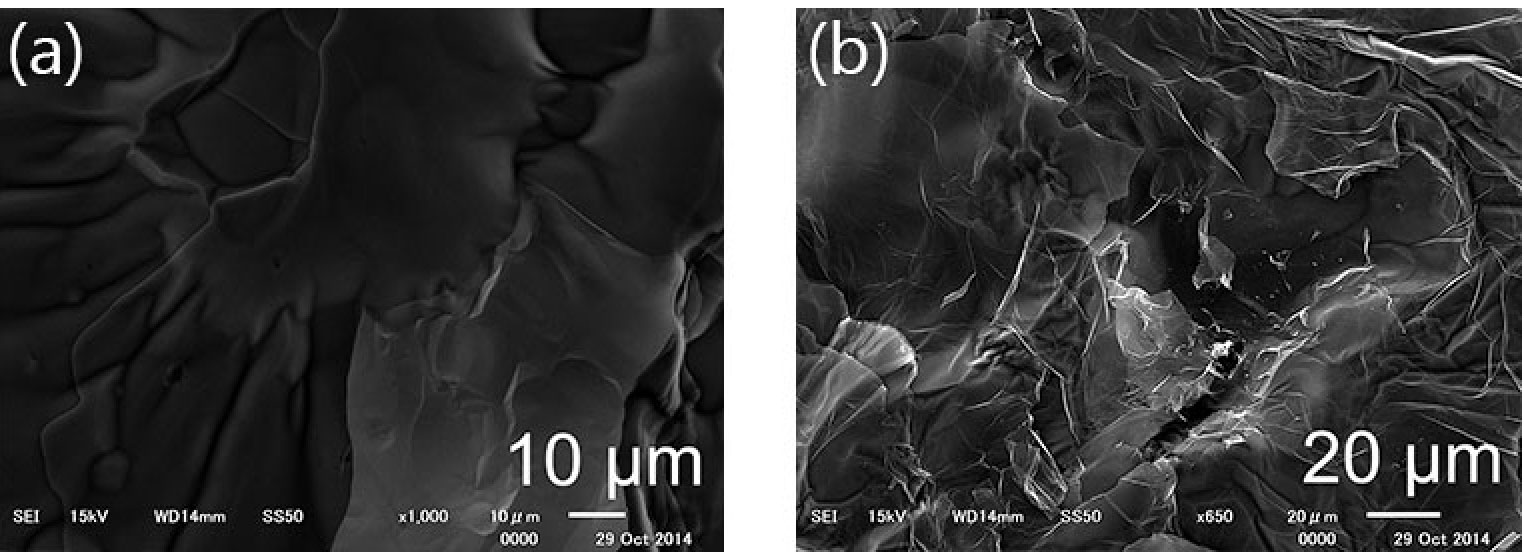}
\caption{Secondary electron SEM images of cleaved surfaces of (a)uPGS-17$\mu$m and (b)uPGS-100$\mu$m.}
\label{SEM-uPGS17cPGS100}
\end{figure}

Figure~\ref{SEM-uPGS17cPGS100}(a) is a secondary electron SEM image of a cleaved surface of uPGS-17$\mu$m.
It was difficult to obtain enough contrast, indicating a quite smooth surface.
A vaguely seen structure of the length scale of  50--100\,$\mu$m here is an alternative looking of the winkles seen by the optical microscope (Fig.~\ref{photo-uPGScPGS}(a)).
The situation is somewhat different for uPGS-100$\mu$m as shown in Fig.~\ref{SEM-uPGS17cPGS100}(b).
The structure is much sharper and smaller in size, and thus the image contrast is much higher. 
Apparently, thicker uPGSs have less uniformity compared to thinner ones because gas is probably more difficult to escape during the carbonization process in the former case.

\begin{figure}[htbp]
\includegraphics[width=0.97\columnwidth]{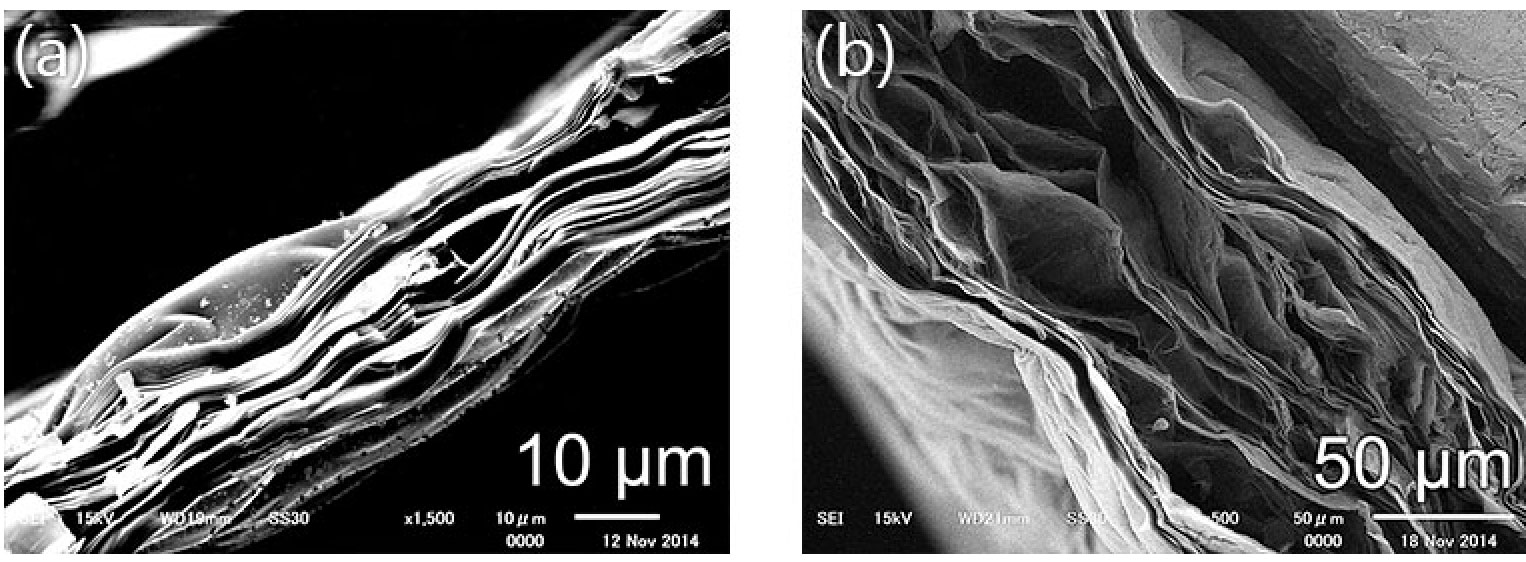}
\caption{Secondary electron SEM images of cross sections of (a)uPGS-17$\mu$m and (b)uPGS-100$\mu$m.}
\label{SEMX-uPGS17cPGS100}
\end{figure}

Internal morphology of uPGS was investigated by cross-sectional secondary-electron SEM imaging (see Figs.~\ref{SEMX-uPGS17cPGS100}(a) and (b)), where the fracture surfaces were obtained by tearing uPGS pieces with a vise.
The images show lamination of wavy layers of sub-$\mu$m thick.  
A similar layer structure has also been observed in the previous TEM studies~\cite{10012555627}. 
Note that each layer is an order of magnitude thinner than raw polymer films before carbonization. 
In uPGS-17$\mu$m the layers are rather densely packed, while in uPGS-100$\mu$m they are loosely packed with large interlayer spaces which may or may not connect to the outside (see a related discussion in Section~4). 
The layer thickness is much thicker than individual graphene sheet.
From the quantum Monte Carlo calculation~\cite{PhysRevLett.102.085303}, the adsorption potential for helium of thin graphite is practically the same as that of bulk graphite if the thickness is larger than eight graphene layers.

\subsection{Observation with scanning tunneling microscope}
\begin{figure}[htbp]
\begin{minipage}{0.48\hsize}
\includegraphics[width=0.98\textwidth]{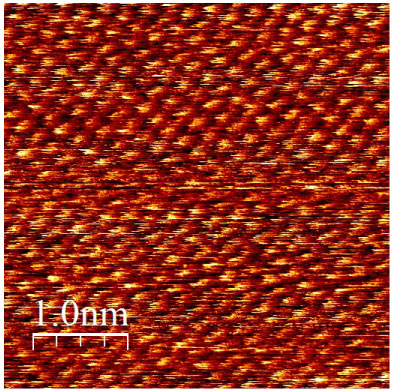}%
\caption{STM topography image of a cleaved surface of uPGS-17$\mu$m. A Pt$_{0.8}$Ir$_{0.2}$ tip was used. The bias voltage $V$ and tunneling current $I$ are 500\,mV and 1.0\,nA, respectively.
\label{stmatom}}%
\end{minipage}
\hfill
\begin{minipage}{0.48\hsize}
\includegraphics[width=0.98\textwidth]{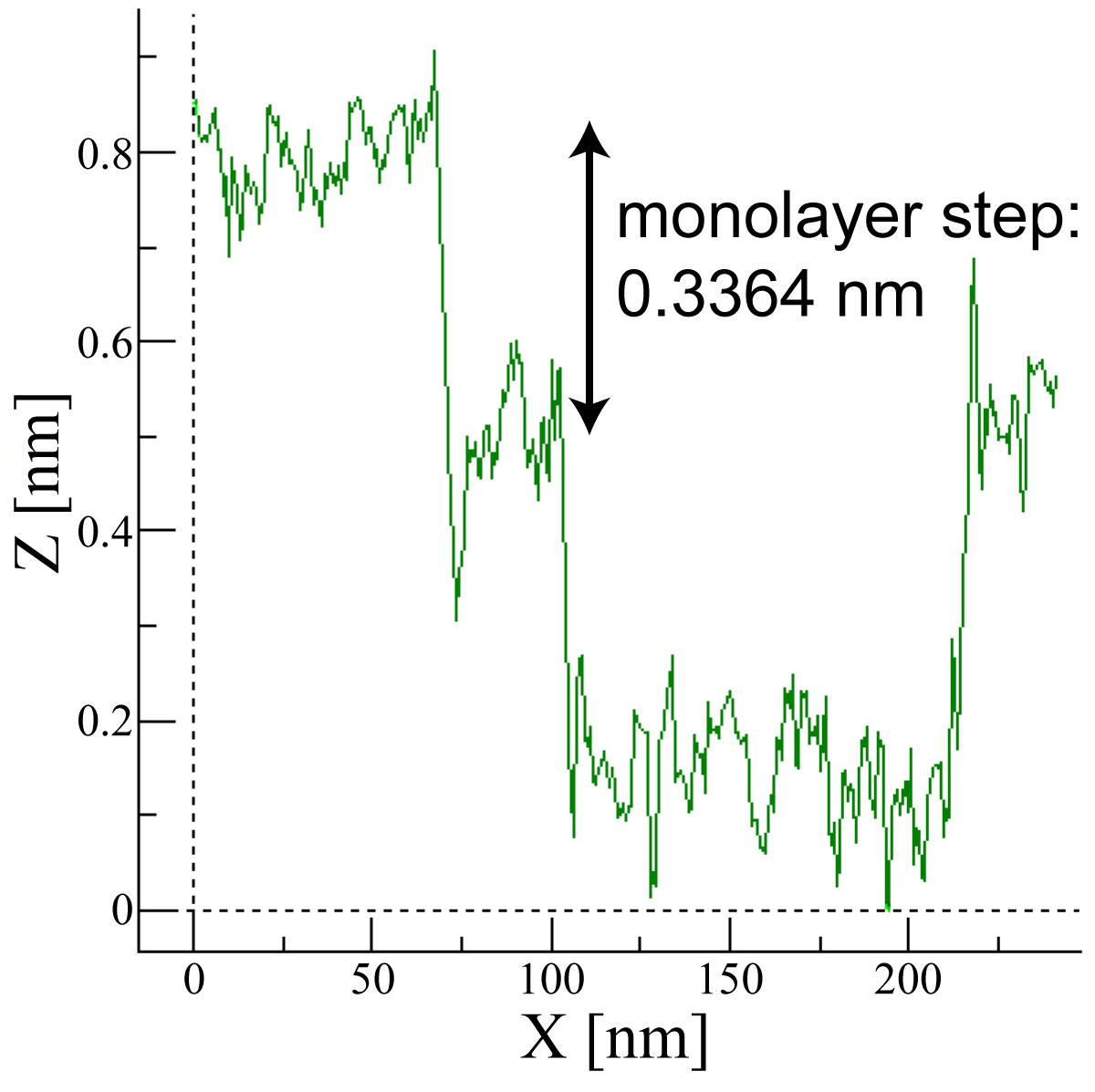}%
\caption{Line profile of an STM image of a cleaved surface of uPGS-17$\mu$m.
\label{stmprofile}}%
\end{minipage}
\end{figure}

The lateral surface structure of uPGS-17$\mu$m was also investigated with a scanning tunneling microscope (STM)~\cite{stm} in atomic scale at room temperature in the air. 
As shown in Fig.~\ref{stmatom}, the honeycomb lattice structure of graphite basal plane is seen almost everywhere on the surface. 
Note that, in STM imaging of a surface of graphite with the normal AB (Bernal) stacking, only carbon atoms belonging to one of two sublattices in topmost graphene are visible.
That is why a triangular lattice rather than the honeycomb lattice is observed in Fig.~\ref{stmatom}.
Besides the wavy structure of the order of 1\,$\mu$m, which would be related to the smaller scale structure seen by SEM, we observed monoatomic height steps with a mean separation of the order of 100\,nm (see Fig.~\ref{stmprofile}). 
The step density is an order of magnitude higher than highly oriented pyrolytic graphite (HOPG)~\cite{PhysRevB.73.085421}.
These steps may affect the critical velocity in superflow experiment for a few layers of $^4$He films to some extent (see Section~5).

\section{Small grains on top surface of PGS}
\begin{figure}[htbp]
\begin{minipage}{0.48\hsize}
\includegraphics[width=0.98\textwidth]{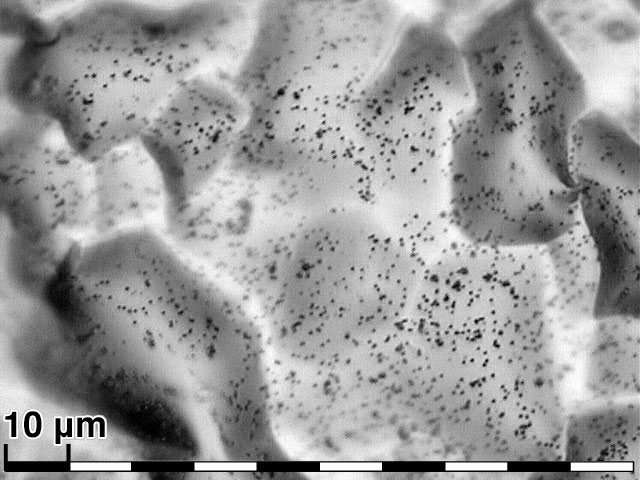}%
\caption{Optical microscope image of a top surface of uncleaved uPGS-10$\mu$m.
\label{photo-uPGS}}%
\end{minipage}
\hfill
\begin{minipage}{0.48\hsize}
\includegraphics[width=0.98\textwidth]{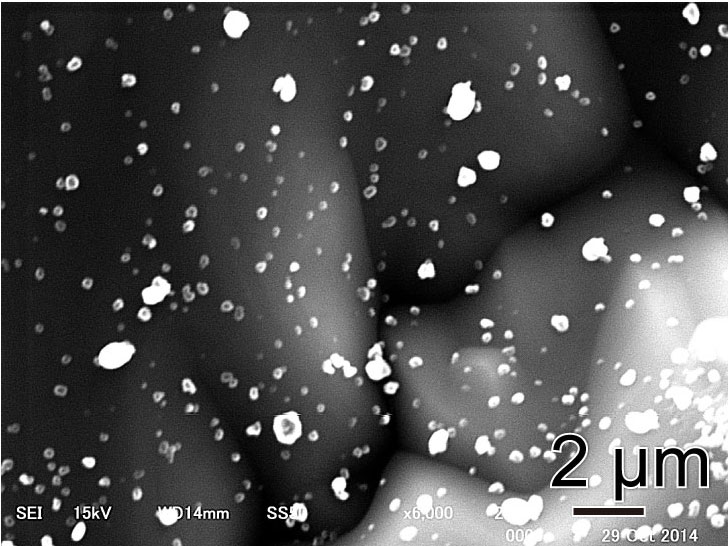}%
\caption{SEM image of a top surface of uncleaved uPGS-17$\mu$m.
\label{uPGS-SEM}}%
\end{minipage}
\end{figure}

We found that $\it{uncleaved}$ top surfaces of uPGS and cPGS are covered by many tiny grains distributing randomly regardless of the sample thickness as shown in Fig.~\ref{photo-uPGS}.
The grain size determined by the secondary electron SEM is from 0.2 to 2\,$\mu$m in diameter, and they cover about 10\% of the top surface in area (see Fig.~\ref{uPGS-SEM}).
The grains are optically black and protruded from the surface judging from back scattered SEM imaging (not shown).
The energy-dispersive X-ray spectroscopy, which is sensitive to boron and heavier elements, detected no traces other than carbon and oxygen. 
They can be removed neither by heating up to 700\,$^{\circ}$C in vacuum ($<10^{-5}$\,mbar) nor by washing with organic solvents or water.
From all these facts including results of Raman spectroscopy measurements shown in Appendix~B, we speculate that the surface grain is a stack of graphite microcrystallites consisting of relatively small numbers of graphene layers.
Since the grains are stuck only on the top surface, they can throughly be removed by surface cleavage with scotch tape as shown in Figs.~\ref{photo-uPGScPGS} and \ref{SEM-uPGS17cPGS100} and possibly with other methods like hydrogen plasma etching.

\section{Adsorption Isotherm Measurement}
Gas adsorption characteristics are crucial to apply a material to substrate for adsorption experiments.
We have made adsorption isotherm measurements of nitrogen molecules (N$_2$) at $T=77$~K for uncleaved uPGS-17$\mu$m.
The sample is a stack of 204 uPGS sheets whose total weight is 1.14\,g.
The size of each sheet is $0.7\times2.1$\,cm$^2$.
The sheets were $\it{loosely}$ packed into a $2.2\times2.2\times0.7$\,cm$^3$ open box made of a thin copper plate. 
They were baked at 700\,$^{\circ}$C in vacuum ($<1.3\times10^{-5}$\,mbar) and then quickly inserted to a vacuum chamber (sample cell) made of brass.
The nominal surface area ($S_{\mathrm{nom}}$) calculated from the outer dimensions of the PGS sheets is 600\,cm$^2$ without taking account of the wavy morphology of the outer surface and the possible interlayer space shown in Fig.~\ref{SEMX-uPGS17cPGS100}(a).

\begin{figure}[bhtp]
\includegraphics[width=0.98\columnwidth]{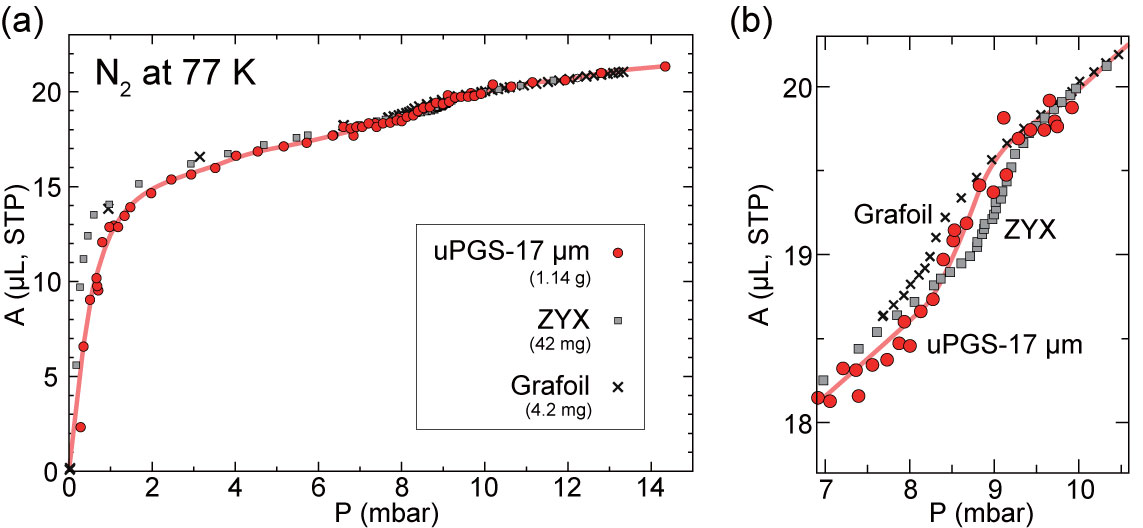}%
\caption{(a) Adsorption isotherms of N$_2$ at $T=77$\,K for uPGS-17$\mu$m (this work; closed circles), ZYX (Ref.~\citenum{LT26nakamura}; squares), and Grafoil (Ref.~\citenum{niimi:4448}; crosses). 
All the $A_{\mathrm{ad}}$ data are scaled to that of uPGS-17$\mu$m at $P\geq12$\,mbar.  
(b) Substeps associated with formation of the $\sqrt{3}\times\sqrt{3}$ commensurate phase of 6.37\,nm$^{-2}$.
\label{plot4_2}}%
\end{figure}

Before discussing the result shown in Fig.~\ref{plot4_2}, let us first explain some technical details to perform this kind of measurement for bulk materials with tiny specific surface area ($s$), i.e., surface area per unit weight.
It is much more difficult than measurements for exfoliated, porous or powder materials. 
In Fig.~\ref{plot4_2} the vertical quantity ($A_{\mathrm{ad}}$) is an amount of N$_2$ adsorbed on the surface, and the horizontal one ($P$) is the equilibrium gas pressure measured by a capacitive pressure gauge~\cite{setra} located at room temperature.
Here an amount of gaseous N$_2$ existing in an open volume (or dead volume) $V_{\mathrm{dead}}$ in the cell has already been subtracted from the raw data. 
$V_{\mathrm{dead}}$ ($\approx 20$\,cm$^3$) was measured precisely with $^4$He gas, which is not adsorbed on the surface at 77\,K, in a separate measurement.
In that measurement, the apparent (skeletal) density of the sample can also be determined from a difference between $V_{\mathrm{dead}}$ values with and without the sample. 
The thermal transpiration effect~\cite{takaishi1963thermal} in the pressure measurement was directly measured for empty cell with N$_2$ and was taken into account at $P \leq$2\,mbar.

The adsorption isotherm data for uPGS-17$\mu$m shown by the closed circles in Fig.~\ref{plot4_2} agree reasonably well with those of Grafoil (crosses)~\cite{niimi:4448} and ZYX (squares)~\cite{LT26nakamura}.
Here, the $A_{\mathrm{ad}}$ data of Grafoil and ZYX are scaled to that of uPGS-17$\mu$m at $P\geq12$\,mbar.
The data for uPGS-17$\mu$m were reproducible among three independent measurements.  
The ``substep'' feature for the $\sqrt{3}\times\sqrt{3}$ commensurate phase formation at $\rho = 6.37$\,atoms/nm$^{2}$~\cite{PhysRevB.13.1446} is seen near $P=9$\,mbar (see Fig.~\ref{plot4_2}(b)).
The small variation of substep structure in different substrates may come from small variations in surface uniformity and/or measuring temperature ($\delta T \approx 0.1$\,K). 
This also assures that the surface of PGS has nearly the perfect graphite structure in atomic scale, i.e., complete graphitization in the production process. 
The surface area ($S$) determined from the upper bound of the substep is 830\,cm$^2$ ($s = 0.073$\,m$^2$/g). 
This value is larger than $S_{\mathrm{nom}}$ by 40\%, which is attributable to the existence of non-negligible internal surface area.
Note that the wavy morphology of outer surface increases $S_{\mathrm{nom}}$ only by 3\%. 
The apparent density of the uPGS-17$\mu$m sample is estimated as 1.76$\pm$0.03\,g/cm$^3$, while that of bulk graphite is 2.26\,g/cm$^3$
This indicates the existence of an internal void space (22\%) not accessible for N$_2$ molecules from outside. 

The adsorption isotherm measurement for uncleaved uPGS-100$\mu$m was even more difficult than uPGS-17$\mu$m because of much smaller $S_{\mathrm{nom}}$, and hence the data show much larger scatterings (not shown).
The apparent density determined from $V_{\mathrm{dead}}$ measurements is 1.20\,g/cm$^3$ for uPGS-100$\mu$m sheets with 26--32 drill holes of 0.5\,mm diameter, which is much smaller than that of uPGS-17$\mu$m, indicating much larger inaccessible intra- and/or inter-sheet void volume. This is in accordance with the cross-sectional secondary-electron SEM images shown in Figs.~\ref{SEMX-uPGS17cPGS100}(a) and (b).
For sheets without the ventilation holes, an even lower value ($=$0.82\,g/cm$^3$) was obtained.

\section{Feasibility Studies of Experiments on Superfluid Thin Films Using uPGS}
In this section, we study the feasibility of AC and DC superflow detection for monolayers of $^4$He adsorbed on a uPGS substrate at temperatures below 1\,K by the torsional oscillator and thermal flow techniques.
Here, among other types, uPGS-10-$\mu$m was chosen as a substrate mainly because of its largest specific surface area and the highest in-plane thermal conductivity~\cite{NAKAMURA2017118}.

In some experiments where so many uPGS sheets have to be used like torsional oscillator experiment, it is not easy to cleave all the surfaces to remove the surface grains discussed in Section~3. 
In that case, one may use uncleaved uPGS.
However, since the grains are well-dispersed point defects rather than line defects like crevices, they would not be so serious obstacles for $^4$He superflow.
The observation of the clear substep in the adsorption isotherm using uncleaved uPGS (Fig.~\ref{plot4_2}) is also encouraging.

\subsection{Torsional oscillator experiment}
The substrate for torsional oscillator experiment will consist of a stack of 1000 uPGS-10-$\mu$m disks of 20\,mm in diameter and silver foils of 2\,$\mu$m in thickness and 8\,mm in diameter.
They will be piled up alternately and be diffusively bonded at 600\,K under pressure to make them a mechanically rigid single piece.
Then it will be tightly packed into a container made of magnesium alloy of 20\,mm in height and 1\,mm wall thickness. 
An estimated total surface area of uPGS is 0.53\,m$^2$.
The silver foils are used not only to assist the thermal conductance of the stack but also to keep proper gaps between the uPGS disks in the outer region as spacers.
Although a small amount of $^4$He may be adsorbed on the silver surface too, it will not affect the superfluid detection as follows.
The adsorption potential of silver for helium is 3 times less than that of graphite~\cite{PhysRevB.46.13967}.
Thus, for example, the second layer of $^4$He adsorbed on uPGS will be in equilibrium with a submonolayer $^4$He adsorbed on silver which will not show superfluidity down to $T = 0$ being tightly bound to the surface.

The superfluidity is detected as a resonance frequency shift ($\Delta f$) of the torsional oscillator due to a change of the moment of inertia ($I$).
Estimated $I$ values of the second layer of $^4$He sample of 7.5\,atoms/nm$^{-2}$, the uPGS disks, and the magnesium-alloy container are $1.8\times10^{-12}$, $3.1\times10^{-7}$, and $2.3\times10^{-7}$\,kg$\cdot$m$^2$, respectively.
$I$ of the silver foils is negligibly small compared to that of the helium film because of the large difference in the diameters. 
Therefore, $\Delta f$ associated with superfluidity of the sample is expected to be 2.9\,mHz at most.
Since a typical accuracy of $\Delta f$ measurement is about 0.02\,mHz, the superfluidity should sufficiently be detected even when the detection efficiency on the uPGS surface was only 0.1.

\subsection{Thermal flow experiments}
An abrupt increase of the thermal conductance along a $^4$He thin film with increasing the coverage at a fixed $T$ or decreasing $T$ at a fixed coverage will provide an alternative evidence for superfluid onset of the film.
For this type of experiment, the substrate crystallinity should be as high as possible not to reduce the superflow.
In addition, the substrate thickness should be as thin as possible not to shunt the temperature gradient by substrate itself when the film is in the normal state.
The experiment by Polanco and Bretz~\cite{POLANCO19801} using a ZYX wafer of $25\times25\times5$\,mm$^3$ was successful to detect superfluid onset of the third layer of $^4$He at $T = 1.0$\,K.
uPGS-10-$\mu$m should potentially be more advantageous for this experiment than ZYX because of its much better crystallinity~\cite{NAKAMURA2017118}.

Next we recall the experiment by Maynard and Chan~\cite{MAYNARD19822090} where a thin HOPG leaf (2$\times$14$\times$0.1\,mm$^3$) was successfully used to detect third sound propagation again in the third layer of $^4$He at $T = 1.0$\,K.
The third sound is a temperature wave (or equivalently an entropy wave) characteristic of superfluid thin films.
They demonstrated that the thermal transport phenomenon of superfluid monolayer can be detected not being shunted by the 100\,$\mu$m-thick graphite substrate.
We thus expect that much thinner uPGS-10$\mu$m will be capable enough to detect a superfluid onset of $^4$He monolayer with an even smaller superfluid density such as the prospected quantum liquid crystal phase in the second layer~\cite{PhysRevB.94.180501}.

\section{Conclusions}
We examined the surface morphology and adsorption isotherm of uncompressed pyrolytic graphite sheet (uPGS), a thin graphite sheet of various nominal thicknesses between 10 and 100\,$\mu$m.
We propose this new type of graphite material as a suitable substrate for experiments to detect superflow in adsorbed monolayers of $^4$He below $T = 1$\,K.
It was found that the cleaved surface of uPGS is quite smooth with the wavy structure of the length scale of several tens $\mu$m from the optical and SEM imaging.
The STM imaging showed that the surface is atomically flat over that length scale with monoatomic  steps roughly every 100\,nm.
We found that many small grains are randomly distributed on PGS surfaces but they are completely absent on cleaved surfaces. 
Using different techniques including Raman spectroscopy, we almost identified them as thin graphite microcrystallites.

On the other hand, Grafoil and ZYX, which are exfoliated graphites commonly used for experiments on two dimensional (2D) $^4$He systems, have quite different surface morphology with much more complicated and shorter length scale structures (platelets) of the order of 10--200\,nm.
Even many deep crevices, which must be created by the exfoliation process, are found on the surface of ZYX.
Our preliminary torsional oscillator measurement eventually suggested the unsuitability of ZYX as an adsorption substrate.

We also made feasibility studies of future torsional oscillator and thermal transport experiments using uPGS-10$\mu$m.
The results show that it is feasible to observe the AC and DC superflow in liquid monolayers or the recently proposed quantum liquid crystal phase in the second layer of $^4$He.
Besides the detection of superfluidity, uPGS will be useful for other applications such as excellent heatsink in a wide temperature range from cryogenic to high temperature, high-conductivity electrode in electrochemistry, etc.

\appendix
\section{Torsional oscillator measurement with ZYX substrate}
Previously, we reported a preliminary attempt to observe superfluidity in $^4$He monolayers adsorbed on ZYX substrate by the torsional oscillator (TO) method~\cite{Kubota2014}.
In that experiment, there were two technical problems.
One was the too large $T$ variation of the resonance frequency of the empty cell, $f_{\mathrm{BG}}(T)$, caused by an accidental application of large strain to the coin silver torsion rod before the measurement.
The other was the instability of $f$, i.e., the too high sensitivity to mechanical shocks and to the 1\,K pot temperature of dilution refrigerator.

We have thus prepared a new experimental setup in which those problems are fixed.
This time, the torsion rod is made of BeCu  (o.d./i.d.\,$= 3.0/1.0$\,mm), and a cell cap is made of 0.5-mm-thick copper. 
The surface area of ZYX substrate is 3.38\,m$^2$. 
The TO cell is fixed onto a massive copper platform (50~mm in diameter, 968\,g in weight) with a copper vibration isolation rod (3~mm in diameter) which is installed on the mixing chamber of dilution refrigerator.  
$f$ and $Q$ values of the empty cell were 1376\,Hz and 2$\times10^{6}$ at $T \leq 4$\,K, respectively. 
A measured $T$ variation of $f$ was less than 2\,mHz at $T \leq 1.6$\,K. 
These performances are equivalent to those realized in the previous TOs~\cite{PhysRevLett.70.3291,1742-6596-150-3-032096, nyeki2017}.

\begin{figure}[htbp]
\centering
\includegraphics[width=0.7\columnwidth]{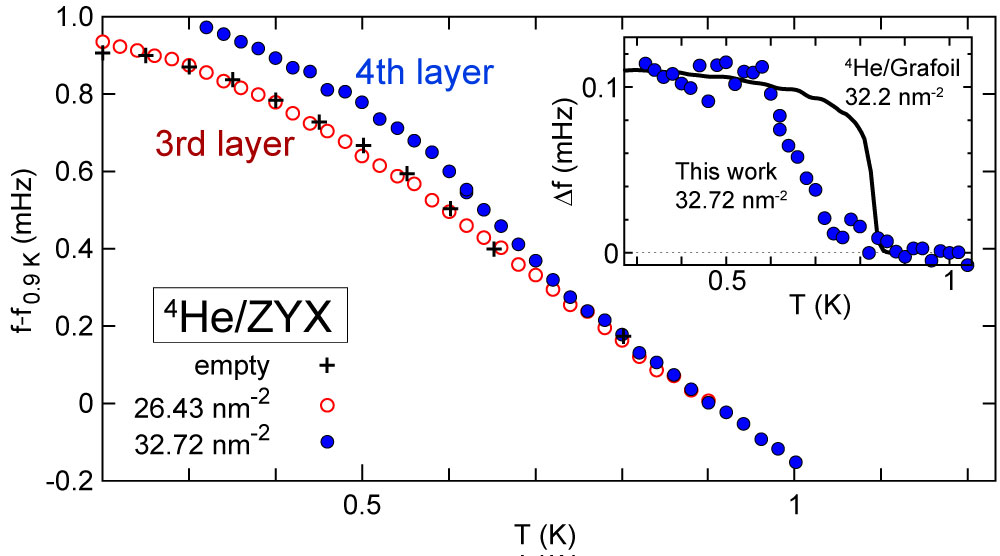}
\caption{Resonance frequency shifts of the torsional oscillator using ZYX substrate with adsorbed $^4$He films of 26.43\,nm$^{-2}$ (open circles) and 32.72\,nm$^{-2}$ (solid circles), and without  $^4$He sample, i.e., the empty cell or background (crosses). Shifts from the values at $T = 0.9$\,K are plotted here. The inset shows frequency shifts for 32.72\,nm$^{-2}$ after subtracting the background shown in the main figure. The line represents data reported in Ref. [\citenum{PhysRevLett.70.3291}] using Grafoil.}
\label{fig_f2}
\end{figure}

With this new setup, we have made preliminary TO measurements for $^4$He thin films of two different densities of 26.43 and 32.72\,nm$^{-2}$ down to 0.2\,K.
The former film consists of a solid bilayer and topmost ``liquid monolayer'', and the latter does a solid bilayer and a topmost ``liquid bilayer'', respectively~\cite{PhysRevLett.70.3291,PhysRevB.47.309}.
At both densities, superfluid responses were detected in the previous TO measurement by Crowell and Reppy (CR)~\cite{PhysRevLett.70.3291} using Grafoil substrate.
In Fig.~\ref{fig_f2}, we plotted measured $f(T)$ with and without the $^4$He films as differences from those at $T = 0.9$\,K.
For 32.72\,nm$^{-2}$ (closed circles) an apparent increase of $f$ from $f_{\mathrm{BG}}(T)$ (crosses) is observed below 0.8\,K, while no meaningful change is seen at 26.43\,nm$^{-2}$.
In the inset, $f(T)$ for 32.72\,nm$^{-2}$ after subtracting $f_{\mathrm{BG}}(T)$ is plotted.
The solid line is the result by CR for a similar density ($= 32.2$\,nm$^{-2}$) where they observed an abrupt jump of $f$ due to 2D superfluidity at an onset temperature $T_{\mathrm{onset}}$ ($= 0.84$\,K) as expected from the Kosterlitz-Thouless (KT) theory~\cite{PhysRevLett.39.1201}.
Here the frequency shift data of CR are normalized so as to be consistent with our data.
On the other hand, our data show a much slower increase below a similar onset temperature.
It is plausible that the crevice structure of ZYX described in the main text weakens the superfluid phase coherence making the transition signature rather extended as in superconducting Josephson networks~\cite{PhysRevB.96.060508}.
This may also explain the absence of superfluid response for 26.43\,nm$^{-2}$ at least down to 0.2\,K.

A jump of superfluid density $\Delta \rho_{\mathrm{s}}$ for our 32.72\,nm$^{-2}$ data is estimated as 0.16$\pm$0.01\,nm$^{-2}$ taking account of the measured mass loading coefficient ($\approx$0.68\,mHz/nm$^{-2}$) and $\Delta f$ ($= 0.11$\,mHz), while 2.2$\pm$0.1\,nm$^{-2}$ is expected from the KT theory with $T_{\mathrm{onset}}=0.84$\,K.
Thus the superfluid detection efficiency ($\eta$) of ZYX is $0.07\pm0.01$ which is a little larger than the $\eta$ value ($= 0.02$--$0.06$) for Grafoil~\cite{PhysRevLett.70.3291,1742-6596-150-3-032096, nyeki2017}. 
Unfortunately, measurements of $Q(T)$ were not stable enough to derive meaningful information.

\section{Raman spectroscopy of uncleaved uPGS surface}
We studied uncleaved uPGS surfaces by a laser Raman microscope~\cite{raman} and compared the results with our previous measurements on cleaved uPGS surfaces~\cite{NAKAMURA2017118}.
The measurements gave us useful insights to define the origin of the surface grains discussed in the main text as a stack of graphite microcrystallites of small numbers of graphene layers.

Raman spectra of graphite and graphene usually have three prominent bands: G, G' and D bands~\cite{PSSB:PSSB201100295}.  
G band (1580\,cm$^{-1}$) is the first order Raman process corresponding to in-plane vibration modes of pairs of sp$^2$ carbon atoms, which is commonly found in aromatic and olefinic molecules~\cite{PhysRevB.61.14095}. 
D band and G' band (also known as D* band or 2D band) are second order ones whose wavenumber vary with the excitation energy. 
D band intensity is an indicator of disorder including edges~\cite{PSSB:PSSB201100295}.
Its overtone, G' band, does not require any disorder and has a strong intensity comparable to that of G band. 
The G' band appears a symmetric peak for monolayer graphene but bilayer graphene has a non-symmetric peak at a slightly higher wavenumber because there are 4 sub-bands due to inter-layer coupling. 
The width and wavenumber increase as we increase the number of graphene layers up to 10~\cite{PSSB:PSSB201100295}. 

\begin{figure}[htbp]
\centering
\includegraphics[width=0.72\columnwidth]{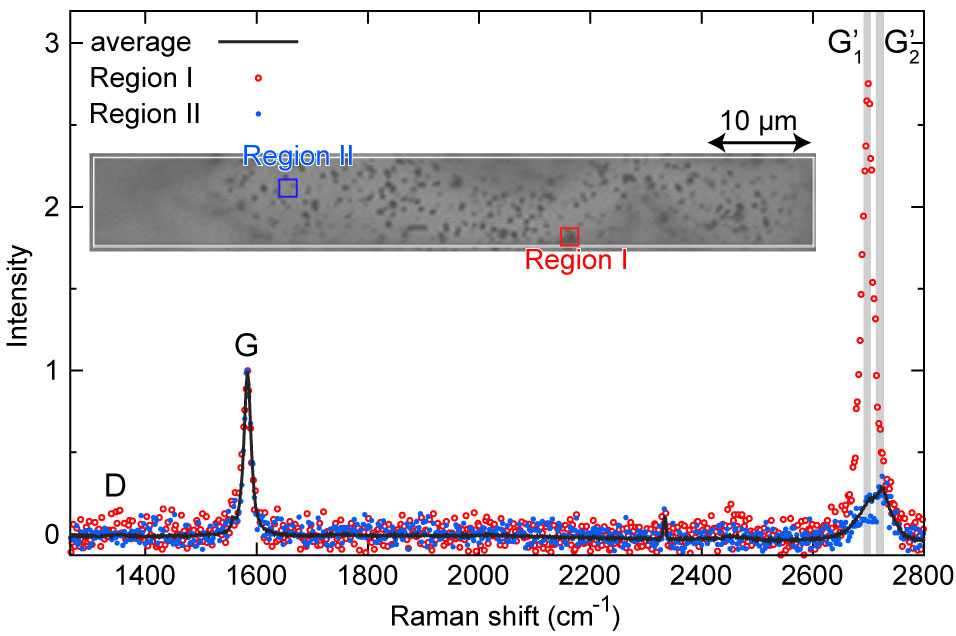}%
\caption{Raman spectra at an uncleaved uPGS-17$\mu$m surface averaged within the region I (open circles) including a large surface grain and the region II (solid circles) including few grains shown in the inset (optical image). The solid line is a spectrum averaged over the whole image.
The intensities are normalized by G peak height. The background intensity of a glass substrate under the uPGS sample has already been subtracted. 
The sharp peak near $2330$\,cm$^{-1}$ comes from the varnish used to fix the uPGS sample to the glass substrate.
\label{spectra}}%
\end{figure}

Figure~\ref{spectra} shows Raman spectra at uncleaved uPGS-17$\mu$m averaged over the following three different areas indicated in the inset (optical microscope image). 
The solid circles represent a spectrum within the region I in the inset including a large grain, and the open circles do that of the region II including few grains. 
The solid line is a spectrum averaged over the whole image.
Here the intensity of each spectrum is normalized by G peak height. 
Apparently, the grains have a strong intensity at lower wavenumbers of G' band (G'$_1$) at $\approx2700$\,cm$^{-1}$. 
This feature is characteristic of (a few layers of) graphene~\cite{PhysRevLett.97.187401}. 
On the other hand, in Region II, the peak appears at slightly higher wavenumber, G'2, which is a common characteristic of bulk graphite.

\begin{figure}[htbp]
\centering
\includegraphics[width=0.72\columnwidth]{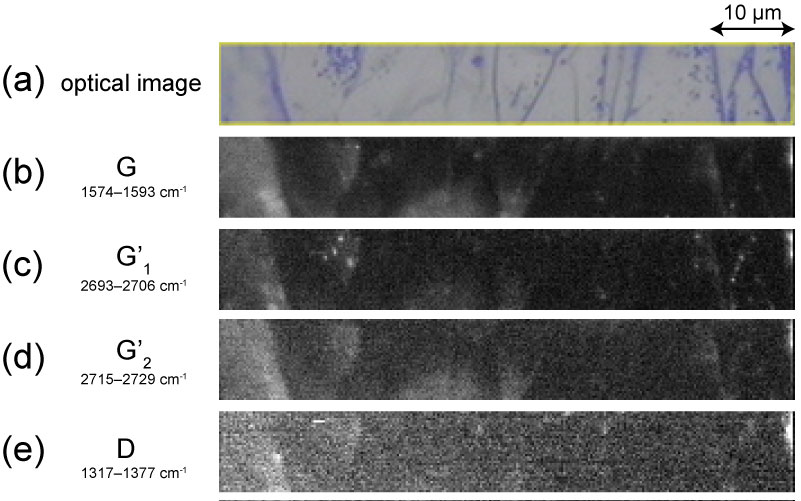}%
\caption{(a)Optical microscope image and (b)--(e)Raman spectroscopy mappings corresponding to G, G'$_{1}$, G'$_{2}$, and D bands at an uncleaved uPGS-100$\mu$m surface. The numbers on the left are spectral windows used for the intensity mappings which are also indicated in Fig.~\ref{spectra}. Brighter contrast means stronger Raman intensity. The area for each image or mapping is 9.5$\times$67.7\,$\mu$m$^2$.
\label{PGS100hi_color}}%
\end{figure}

Spectral mappings for an uncleaved uPGS-100$\mu$m surface shown in Fig.~13 provide us more information.  
While the G'$_2$ mapping (Fig.~\ref{PGS100hi_color}(d), indicator of graphite) only has a shallow contrast due to the defocusing, G'$_1$ mapping (Fig.~\ref{PGS100hi_color}(c), indicator of graphene) has many spots with strong intensity. 
These spots correspond to the grains in the optical image, Fig.~\ref{PGS100hi_color}(a). 
G band mapping, Fig.~\ref{PGS100hi_color}(b), also show similar spots. 
However, in those spots, Raman intensities at other wavenumbers, not only in G band, are equally enhanced. 
Thus the G band distribution does not indicate difference in the material composition. 

At uncleaved uPGS surface we always observed a finite D band intensity, which is sensitive to symmetry breaking defects or edges~\cite{rohtua}, while it is not detectable at cleaved surface~\cite{NAKAMURA2017118}(see also Table~\ref{ratio_2}).
However, it does not seem to be directly related to the surface grains, since 
one cannot find a strong correlation in a D band mapping (Fig.~\ref{PGS100hi_color}(e)) with the grain distribution in the optical image (Fig.~\ref{PGS100hi_color}(a)).
The defects may be located just below the top surface~\cite{ncomms5224}.

\begin{table}[htbp]
\begin{center}
\caption{Intensity ratios of {$I$(G)/$I$(G') and $I$(D)/$I$(G)} in Raman spectra for uncleaved {and cleaved} surfaces of uPGS.}\label{ratio_2}
\begin{tabular}{rcccc}
\hline \hline
material & nominal thickness & $I$(G)/$I$(G')  &  $I$(D)/$I$(G) &  reference\\
\hline
uncleaved uPGS & $100~\mu$m  & {1.3--3.3} &{0.028(5)}  &{this work}\\
{uncleaved uPGS} & $25~\mu$m &{0.3--3.0}  & {0.018(11)}  &{this work}\\
{uncleaved uPGS} & $17~\mu$m  & {0.3--3.2} & {0.013(5)}  &{this work}\\
{cleaved uPGS} &{$100~\mu$m} & {3.2(1)}  &{$<10^{-3}$} & {Ref.~\citenum{NAKAMURA2017118}}\\
{cleaved uPGS} & {$25~\mu$m} & {3.1(1)}  & {$<10^{-3}$} & {Ref.~\citenum{NAKAMURA2017118}}\\
{cleaved uPGS} & {$17~\mu$m} & {3.1(1)}  & {$<10^{-3}$} & {Ref.~\citenum{NAKAMURA2017118}}\\
\hline \hline
\end{tabular}
\end{center}
\end{table}

\begin{acknowledgements}
We are grateful to Yoshiya Sakaguchi, Hiroyuki Hase, and Makoto Nagashima of Automotive \& Industrial Systems Company of Panasonic corporation for providing us the uPGS and cPGS samples. 
We also thank Hideki Sato for his assistance in the STM measurement and Takenori Fujii for his contribution in the energy-dispersive X-ray spectroscopy measurement.  
The laser Raman microscope used in this work was supplied by MERIT program, The University of Tokyo.
\end{acknowledgements}


%
%

\end{document}